\documentclass[a4paper,12pt]{article}
\usepackage{amsmath,amsfonts,amssymb,amscd,enumerate}

\def\C{\mathbb{C}}

\def\R{\mathbb{R}}

\def\p{\partial}

\def\F{\mathcal{F}}
\def\G{\mathcal{G}}
\def\H{\mathcal{H}}
\def\x{\hat x}

\begin{document}
\begin{center}
{\Large \bf Covariant realizations of kappa-deformed  space}
\end{center}
\bigskip

\begin{center}
 Stjepan Meljanac ${}^{a,}$\footnote{e-mail: meljanac@thphys.irb.hr},
 Sa\v{s}a  Kre\v{s}i\'{c}-Juri\'{c} ${}^{b,}$\footnote{e-mail: skresic@fesb.hr}
 and Marko Stoji\'c ${}^{a,}$\footnote{e-mail: marko.stojic@zg.htnet.hr}\\
\end{center}

\begin{center}
{${}^a$ Rudjer Bo\v{s}kovi\'c Institute, Bijeni\v cka  c. 54, 10002 Zagreb, Croatia}
\end{center}

\begin{center}
{${}^b$ Faculty of Natural and Mathematical Sciences, Teslina 12, 21000 Split, Croatia}
\end{center}

\bigskip

\setcounter{page}{1}
\bigskip

\begin{abstract}

We study a Lie algebra type $\kappa$-deformed space with undeformed rotation algebra and commutative
vector-like Dirac derivatives in a covariant way. Space deformation depends on an arbitrary vector. Infinitely
many covariant realizations in terms of commuting coordinates of undeformed space and their derivatives
are constructed. The corresponding coproducts and star products are found and related in a new way.
All covariant realizations are physically equivalent. Specially, a few simple realizations are found
and discussed. The scalar fields, invariants and the notion of invariant integration is discussed
in the natural realization.

\end{abstract}

 PACS number(s):


\newpage

\section{Introduction}

In the last decade there has been a growing interest in the formulation of physical theories defined on noncommutative (NC) spaces.
Consistency of such theories and their implications were studied in \cite{1}-\cite{5}. It is important to classify NC spaces and investigate
their properties, and particularly to develop a unifying approach to a generalized theory for such spaces that are convenient
for physical applications. The notion of generalized symmetries and their role in the analysis of NC spaces is also crucial. In
order to make a contribution in this direction, we analyze a NC space of the Lie algebra type, in particular the so-called $\kappa$-deformed
space introduced in \cite{6}-\cite{8}.

For simplicity we restrict our attention to $\kappa$-deformed Euclidean space, although the analysis can  be easily extended to
$\kappa$-deformed Minkowski space. The noncommutativity of coordinates depends on a deformation parameter which is an arbitrary
vector $a\in \R^n$. The dimensional parameter $|a|=1/\kappa$ has a very small length which yields the undeformed Euclidean space
in the limit as $|a|\to 0$. The NC coordinates and the generators of generalized rotations form an extended Lie algebra. The subalgebra
formed by the rotation generators is the ordinary $SO_a(n)$ algebra, i.e. the ordinary Lorentz algebra in the case of
$\kappa$-deformed Minkowski space. Dirac derivatives are assumed to commute mutually and transform as a vector representation under
$SO_a(n)$ algebra. This $\kappa$-deformed space was studied by different groups, from both the mathematical and physical point
of view \cite{9}-\cite{27}. There is also an interesting connection to the Doubly Special Relativity program \cite{15}, \cite{16}.
Realizations of NC spaces in terms of commutative coordinates and derivative operators were obtained and discussed in the cases
of symmetric ordering and normal (left and right) ordering of NC coordinates \cite{13}, \cite{21}. An infinite family of noncovariant
realizations was found in \cite{22}. Although a single space may be realized in many different ways, physical results do not
depend on concrete realizations, i.e. orderings (see also \cite{24}).

Our motivation in this paper is to construct covariant realizations for general $\kappa$-deformed Euclidean spaces depending
on an arbitrary deformation vector $a$. We analyze such spaces by using the methods developed for deformed single and multimode oscillators
in the Fock space representations \cite{29}-\cite{36}. In particular, we use the methods for constructing deformed creation and annihilation
operators in terms of ordinary bosonic multimode oscillators, i.e. a kind of bosonization \cite{29}, \cite{30}, \cite{35}. Also, we employ the
construction of transition number operators and, in general, of generators proposed in \cite{30}, \cite{31}, \cite{34}.

The simple connection between the creation and annihilation operators with NC coordinates and Dirac derivatives is established
by using the Bargman type representation. We find infinitely many new covariant realizations in terms of commutative coordinates
and derivative operators. The realizations depend on certain parameter functions, but they are all on an equal footing and the
physical results do not depend on them. For a special choice of the parameter functions we obtain some
particularly simple realizations: covariant left, right and natural realizations. These realizations are considered in
detail, and a coproduct and star product are associated to each of them.

The outline of the paper is as follows. In Sect. 2 we introduce a Lie algebra type of $\kappa$-deformed Euclidean space. We also
define the rotation algebra $SO_a(n)$ which is compatible with $\kappa$-deformations, and introduce the Dirac derivative and the
Laplace operator. The shift operator which plays an important role in describing the algebra generated by NC coordinates,
the rotation generators and the Dirac derivative are also introduced. Special consideration is given to the generalized Leibniz's
rule and coproduct. We derive expressions for the coproduct of the rotation generators and the Dirac derivative, and of the left
and right deformations of the ordinary derivative. Sect. 3 deals with covariant realizations of the $\kappa$-deformed Euclidean
space and the operators introduced in Sect. 2. We find an infinite family of covariant realizations in terms of commutative
coordinates and derivative operators. In particular, we obtain two types of realizations (type I and type II) which depend
on an arbitrary parameter function $\varphi$. For special choices of $\varphi$, we construct particularly simple covariant realizations:
left, right, symmetric and natural realizations. In Sect. 4 we consider the star product for the realizations discussed in Sect. 3.
A general formula for the star product is given to second order in the deformation parameter $a$, and closed form expressions
are obtained in the left, right and symmetric realizations. We also introduce the notion of equivalent star products using
similarity transformations. We show that any star product in the realization of type I can be obtained from the star product
in the right realization using  similarity transformations. In subsection 4.4 we introduce scalar fields in NC coordinates
and demonstrate their simple properties in terms of the natural realization. Constructions of invariants and invariant
integration on NC spaces are also presented. Finally, in Sect. 5 a short conclusion is given.

\section{Kappa-deformed Euclidean space}

Consider a Lie algebra type noncommutative (NC) space generated by coordinates $\x_1, \x_2,\ldots ,\x_n$ satisfying
the commutation relations
\begin{equation}\label{2.1}
[\x_{\mu},\x_{\nu}]=i(a_{\mu}\x_{\nu}-a_{\nu}\x_{\mu}),
\end{equation}
where $\mu,\nu=1,\dots,n$ and $a_1,a_2,\dots ,a_n$ are componenets of a vector $a\in \R^n$ which describes a
deformation of the Euclidean space \cite{19}-\cite{22}. The structure constants are given by
\begin{equation}\label{2.2}
C_{\mu\nu\lambda}=a_{\mu}\delta_{\nu\lambda}-a_{\nu}\delta_{\mu\lambda}.
\end{equation}
In the limit as $a\to 0$, we have $\x_{\mu}\to x_{\mu}$, the ordinary commutative coordinates.

Let $SO_a(n)$ be the ordinary rotation algebra with generators $M_{\mu\nu}$ satisfying
\begin{align}
M_{\mu \nu} &= -M_{\nu \mu}, \label{2.3A}  \\
[M_{\mu\nu},M_{\lambda\rho}] &= \delta_{\nu\lambda}M_{\mu\rho}-\delta_{\mu\lambda}M_{\nu\rho}
-\delta_{\nu\rho}M_{\mu\lambda}+\delta_{\mu\rho}M_{\nu\lambda}. \label{2.3B}
\end{align}
We require that the rotation generators $M_{\mu\nu}$ and the coordinates $\x_{\mu}$ form an extended Lie algebra. The most
general form of the commutator $[M_{\mu\nu},\x_{\lambda}]$ must be linear in $M_{\mu\nu}$ and $\x_{\lambda}$,
antisymmetric in the indices $\mu$ and $\nu$, and with the smooth limit $[M_{\mu\nu},\x_{\lambda}]\to
x_{\mu}\delta_{\nu\lambda}-x_{\nu}\delta_{\mu\lambda}$ as $a\to 0$. The required form is given by
\begin{equation}
\begin{split}
[M_{\mu\nu},\x_{\lambda}] &=\x_{\mu}\, \delta_{\nu\lambda}-\x_{\nu}\, \delta_{\mu\lambda}\\
&+is\, a_{\lambda}\, M_{\mu\nu}-it\, \left(a_{\mu}\, M_{\nu\lambda}-a_{\nu}\, M_{\mu\lambda}\right)+iu\, a_{\alpha}\,
\left(M_{\alpha\mu}\, \delta_{\nu\lambda}-M_{\alpha\nu}\, \delta_{\mu\lambda}\right)
\end{split}
\end{equation}
for some $s,t,u \in \R$, where summation over repeated indices in assumed. The necessary and sufficient condition for
consistency of the extended Lie algebra is that the Jacobi identity holds for all combinations of the generators
$M_{\mu\nu}$ and $\x_{\lambda}$. One can verify that this is satisfied for the unique values of the parameters
$s=u=0$ and $t=1$ \cite{22}. Hence,
\begin{equation}\label{2.15}
[M_{\mu\nu},\x_{\lambda}]=\x_{\mu}\, \delta_{\nu\lambda}-\x_{\nu}\, \delta_{\mu\lambda}-i\left(a_{\mu}\,
M_{\nu\lambda}-a_{\nu}\, M_{\mu\lambda}\right).
\end{equation}

Having introduced the rotation algebra $SO_a(n)$, it is natural to consider the Dirac derivatives $D_{\mu}$ satisfying
\begin{align}
[D_{\mu},D_{\nu}]&=0, \label{2.24} \\
[M_{\mu\nu},D_{\lambda}]&= \delta_{\nu\lambda}\, D_{\mu}-\delta_{\mu\lambda}\, D_{\nu}. \label{2.25}
\end{align}
The generators $M_{\mu}$ and $D_{\lambda}$ form the undeformed $ISO_a(n)$ algebra, i.e. Poincar\'{e} algebra in the case of 
$\kappa$-deformed Minkowski space. Note that the operator $D^2=D_{\mu}D_{\mu}$ is
invariant under rotations since $[M_{\mu\nu},D^2]=0$.

We also wish to define commutation relations for $D_{\mu}$ and $\x_{\nu}$. The consistency condition is that the
Jacobi identity is satisfied for all combinations of the generators $M_{\mu\nu}$, $D_{\lambda}$ and $\x_{\rho}$.
It can be shown \cite{22} that the correct form of the commutator is given by
\begin{equation}\label{2.4}
[D_{\mu},\x_{\nu}] = \delta_{\mu\nu}\, \sqrt{1-a^2\, D^2}+iC_{\mu\alpha\nu}\, D_{\alpha}.
\end{equation}
The algebra generated by $D_{\mu}$ and $\x_{\nu}$ is a deformed Heisenberg algebra. We note that, in the limit as $a\to 0$,
the commutation relations \eqref{2.1}, \eqref{2.24} and \eqref{2.4} yield the ordinary undeformed Heisenberg algebra.
Hence, $D_{\mu}\to \p_{\mu}$ and $\hat{x}_\mu \to x_\mu$  as $a\to 0$, where $\p_{\mu}=\frac{\p}{\p x_{\mu}}$.
Particularly, in the one-dimensional case $n=1$, Eq. \eqref{2.4} leads to a generalized uncertainty relation with
mininal length \cite{37}.

A function $f(\x,D)$, where $f(\x,D)$ denotes a formal power series in monomials $\x_{\mu_1}\x_{\mu_2}\ldots
\x_{\mu_n}$ and $D_{\nu_1} D_{\nu_2} \ldots D_{\nu_m}$, is said to be $SO_a(n)$-invariant if
\begin{equation}
[M_{\mu\nu},f(\x,D)] = 0 \quad \mbox{for all}\quad \mu, \nu.
\end{equation}
We introduce the $SO_a(n)$-invariant Laplace operator $\square$ satisfying the commutation relations
\begin{align}
[M_{\mu\nu},\square] &= 0, \label{2.5} \\
[\square, \x_{\mu}] &= 2D_{\mu}.  \label{2.6}
\end{align}
The Laplace operator can be expressed in terms of the operator $D^2$. Let us assume the Ansatz $\square = F(D^2)$, where $F$
is analytic, and impose the boundary condition $F(D^2)\to \p^2$ as $a\to 0$. Then Eq. \eqref{2.5} is automatically satisfied,
and Eqs. \eqref{2.4} and \eqref{2.6} together with the boundary condition yield
\begin{equation}\label{2.7}
\square = \frac{2}{a^2}\left(1-\sqrt{1-a^2 D^2}\right).
\end{equation}
It follows from here that
\begin{equation}
D^2 = \square \left(1-\frac{a^2}{4}\square\right).
\end{equation}

\subsection{The shift operator}

At this point it is convenient to introduce the shift operator $Z$ via the commutation relations
\begin{align}
[Z,\x_{\mu}] &= ia_{\mu} Z, \\
[Z,D_{\mu}] &= 0.
\end{align}
We assume that $Z\to 1$ in the limit as $a\to 0$. The shift operator acts on an arbitrary function $f(\x)$ by
\begin{equation}
Zf(\x) = f(\x+ia) Z.
\end{equation}
We note that the inverse shift operator $Z^{-1}$ satisfies the same relations as $Z$ when $a_{\mu}$ is replaced by
$-a_{\mu}$. The shift operator can be expressed in terms of the Dirac
derivative $D$ as follows. Using Eqs. \eqref{2.4}
and \eqref{2.6} one can verify that
\begin{equation}\label{add-1}
\Big[-iaD-\frac{a^2}{2}\square, \x_{\mu}\Big] = -ia_{\mu}\left(-iaD+\sqrt{1-a^2 D^2}\right),
\end{equation}
where $aD=a_{\mu} D_{\mu}$. Inserting Eq. \eqref{2.7} for the Laplace operator into Eq. \eqref{add-1}, we obtain
$[Z^{-1},\x_{\mu}]=-ia_{\mu} Z^{-1}$, where
\begin{equation}\label{2.23}
Z^{-1}=-iaD+\sqrt{1-a^2 D^2}.
\end{equation}
Inverting the above expression, we find
\begin{equation}\label{2.12}
Z=\frac{iaD+\sqrt{1-a^2 D^2}}{1-a^2 D^2 +(aD)^2}.
\end{equation}

It is interesting to note that the algebra generated by $\x_{\mu}$, $M_{\mu\nu}$ and $D_{\mu}$ can be described using the shift
operator $Z$ as
\begin{align}
\x_{\mu} Z \x_{\nu} &= x_{\nu} Z \x_{\mu}, \\
[D_{\mu},\x_{\nu}] &= \delta_{\mu\nu} Z^{-1}+ia_{\mu} D_{\nu}.
\end{align}
The remaining commutation relations for $[M_{\mu\nu},M_{\lambda\rho}]$, $[M_{\mu\nu},\x_{\lambda}]$ and $[M_{\mu\nu},D_{\lambda}]$
are satisfied by representing $M_{\mu\nu}$ in a unique way as
\begin{equation}\label{2.17}
M_{\mu\nu} = (\x_{\mu} D_{\nu}-\x_{\nu} D_{\mu}) Z.
\end{equation}
We will justify this relation later when we consider the so-called natural realization in Sect. \ref{covariant-realizations}.

\subsection{The Leibniz rule and coproduct}
\label{coproduct}

Now we turn our attention to the generalized Leibniz rule and coproduct. The commutator of $M_{\mu\nu}$ with an
arbitrary function $f(\x)$ is given by
\begin{equation}\label{Mf}
\begin{split}
[M_{\mu\nu},f] &= (M_{\mu\nu}f)\, \\
&+ ia_{\mu} \left[\left(D_{\lambda}-\frac{ia_{\lambda}}{2} \square\right)\, Zf\right] M_{\lambda\nu}
-ia_{\nu}\left[\left(D_{\lambda}-\frac{ia_{\lambda}}{2}\square\right)\, Zf\right] M_{\lambda\mu}.
\end{split}
\end{equation}
This relation can be shown by using Eq. \eqref{2.15} and proceeding by induction on the degree of monomials in
$\x_{\mu}$. From Eq. \eqref{Mf} we obtain the coproduct for $M_{\mu\nu}$,
\begin{equation}
\begin{split}
\triangle M_{\mu\nu} &= M_{\mu\nu}\otimes \mathbf{1}+\mathbf{1}\otimes M_{\mu\nu} \\
&+ia_{\mu}\left(D_{\lambda}-\frac{ia_{\lambda}}{2}\square\right)\, Z\otimes
M_{\lambda\nu}-ia_{\nu}\left(D_{\lambda}-\frac{ia_{\lambda}}{2}\square\right)\, Z\otimes M_{\lambda\mu}.
\end{split}
\end{equation}
Similarly, one can show that the commutator of $D_{\mu}$ with $f(\x)$ is given by
\begin{equation}
[D_{\mu},f] = (D_{\mu}f) Z^{-1}+ia_{\mu} (D_{\lambda} Zf) D_{\lambda}-\frac{ia_{\mu}}{2} (\square\, Zf)\, iaD,
\end{equation}
hence for the coproduct we have
\begin{equation}\label{coproductD}
\triangle D_{\mu} = D_{\mu}\otimes Z^{-1}+\mathbf{1}\otimes D_{\mu}+ia_{\mu} (D_{\lambda} Z)\otimes
D_{\lambda}-\frac{ia_{\mu}}{2} \square\, Z\otimes iaD.
\end{equation}
Furthermore, the coproduct for the shift operator $Z$ is simply
\begin{equation}
\triangle Z = Z\otimes Z.
\end{equation}
Some examples of the Poincar\'{e} invariant interpretation of NC spaces and of the twisted
Poincar{\'e} coalgebra were also considered in \cite{38}-\cite{40}.

It is interesting to note that the operators $D_{\mu}$, $\square$ and $Z$ can be expressed in terms of auxiliary derivatives
$\p_{\mu}^L$ and $\p_{\mu}^R$ satisfying the following commutation relations:
\begin{align}
[\p_{\mu}^L, \p_{\nu}^L] &= 0, \label{2.10A} \\
[\p_{\mu}^L, \x_{\nu}] &= \delta_{\mu\nu}\, Z^{-1}, \label{2.10B}
\intertext{and}
[\p_{\mu}^R, \p_{\nu}^R] &= 0, \label{2.11A} \\
[\p_{\mu}^R, \x_{\nu}] &= \delta_{\mu\nu}+ia_{\nu} \p_{\mu}^R.  \label{2.11B}
\end{align}
Equations \eqref{2.10A}-\eqref{2.10B} and \eqref{2.11A}-\eqref{2.11B} are consistent with the commutation relation
\eqref{2.1}, and hence both sets of equations define a deformed Heisenberg algebra. One can think of $\p_{\mu}^L$ and
$\p_{\mu}^R$ as ``left'' and ``right'' deformations of the ordinary derivative $\p_{\mu}$. Indeed, using the
commutation relations \eqref{2.10B} and \eqref{2.11B} one can show that the coproducts of $\p_{\mu}^L$ and $\p_{\mu}^R$
are given by
\begin{align}
\triangle \p_{\mu}^L &= \p_{\mu}^L \otimes Z^{-1}+\mathbf{1}\otimes \p_{\mu}^L, \label{coproductL} \\
\triangle \p_{\mu}^R &= \p_{\mu}^R \otimes \mathbf{1}+Z\otimes \p_{\mu}^R \label{coproductR},
\end{align}
and $\p_{\mu}^R =\p_{\mu}^L Z.$
Hence, the coproducts $\triangle \p_{\mu}^L$ and $\triangle \p_{\mu}^R$ are left and right deformations of the ordinary coproduct $\triangle
\p_{\mu} = \p_{\mu}\otimes \mathbf{1}+\mathbf{1}\otimes \p_{\mu}$, respectively. In fact, the coproducts $\triangle \p_{\mu}^L$
and $\triangle \p_{\mu}^R$ are given by Eqs. \eqref{coproductL}-\eqref{coproductR} if and only if
Eqs. \eqref{2.10B} and \eqref{2.11B} hold. One can show that the operators $D_{\mu}$, $\square$ and $Z$ are expressed
in terms of the left and right deformation derivatives as
\begin{align}
D_{\mu} &= \p_{\mu}^L+\frac{ia_{\mu}}{2} \square, \label{Dleft} \\
\square &= (\p^L)^2\, Z, \\
Z &= 1+i(a\p^L)\, Z = \frac{1}{1- ia\p^L}\\
\intertext{and}
D_{\mu} &= \p_{\mu}^R\, Z^{-1}+\frac{ia_{\mu}}{2} \square, \label{Dright} \\
\square &= (\p^R)^2\, Z^{-1}, \\
Z &= 1+ia\p^R.
\end{align}

The algebra generated by $\x_{\mu}$, $M_{\mu\nu}$ and $D_{\mu}$ is covariant under the action of the rotation group
$SO(n)$. Indeed, let $R\in SO(n)$ be a rotation matrix and denote the transformed variables by $\x_{\mu}^\prime=R_{\mu\alpha}\x_{\alpha}$,
$D_{\mu}^\prime=R_{\mu\alpha} D_{\alpha}$, $M_{\mu\nu}^\prime = R_{\mu\alpha} R_{\nu\beta} M_{\alpha\beta}$ and
$a_{\mu}^\prime = R_{\mu\alpha} a_{\alpha}$. Then Eqs. \eqref{2.7} and \eqref{2.12} immediately yield
\begin{equation}
\square^\prime = \square \quad \mbox{and}\quad Z^\prime = Z,
\end{equation}
and the transformed generators $\x_{\mu}^\prime$, $M_{\mu\nu}^\prime$ and $D_{\mu}^\prime$ satisfy the relations
\begin{align}
\x_{\mu}^\prime Z \x_{\nu}^\prime &= \x_{\nu}^\prime Z \x_{\mu}^\prime, \\
[D_{\mu}^\prime, \x_{\nu}^\prime] &= \delta_{\mu\nu} Z^{-1} + i a_{\mu}^\prime\, D_{\nu}^\prime, \\
M_{\mu\nu}^\prime &= (\x_{\mu}^\prime D_{\nu}^\prime-\x_{\nu}^\prime D_{\mu}^\prime) Z.
\end{align}

\section{Covariant realizations}
\label{covariant-realizations}

A realization of the NC coordinates $\x_{\mu}$ in terms of ordinary commutative coordinates and their derivatives
was found using the Bargman representation and the methods developed in \cite{29}, \cite{30}, \cite{34}, \cite{35}.
The goal of this section is to find covariant realizations of the algebra generated by the NC coordinates $\x_{\mu}$, the rotation
generators $M_{\mu\nu}$ and the Dirac derivatives $D_{\mu}$. The realizations are found in terms of functions of ordinary
coordinates $x_1,x_2,\ldots ,x_n$ and their derivatives $\p_1,\p_2,\ldots ,\p_n$ which generate the Heisenberg algebra
$[x_{\mu},x_{\nu}]=[\p_{\mu},\p_{\nu}]=0$ and $[\p_{\mu},x_{\nu}]=\delta_{\mu\nu}$. In general, these functions will satisfy
a system of coupled partial differential equations (PDE) determined by the commutation relations for $\x_{\mu}$, $M_{\mu\nu}$ and $D_{\mu}$.
In the following we derive such systems of PDE's and then we consider their solutions.

The most general Ansatz for $\x_{\mu}$ is
\begin{equation}\label{3.6}
\x_{\mu} = x_{\alpha} \Phi_{\alpha\mu}(A,B),
\end{equation}
where $\Phi_{\alpha\mu}$ is a function of the commuting variables $A=ia\p$ and $B=a^2 \p^2$, and satisfies the boundary condition
$\Phi_{\alpha\mu}(0,0)=\delta_{\alpha\mu}$. This realization is covariant under the orthogonal transformation
$R \in SO(n)$ (c.f. Sect. \ref{coproduct} and $x_{\mu}^\prime =R_{\mu\alpha}x_{\alpha}$, $\p_{\nu}^\prime=R_{\mu\alpha}\p_{\alpha}$),
i.e. under the action of the generators
\begin{equation}
M_{\mu\nu}^0 = x_{\mu}\p_{\nu}-x_{\nu}\p_{\mu}+a_{\mu}\frac{\p}{\p a_{\nu}}-a_{\nu}\frac{\p}{\p a_{\mu}}.
\end{equation}
We consider a particular form of the above Ansatz given by
\begin{equation}\label{3.1}
\x_{\mu} = x_{\mu}\varphi + i(ax)\left(\p_{\mu}\, \beta_1+ia_{\mu}\p^2\, \beta_2\right)+i(x\p)
\left(a_{\mu}\gamma_1+ia^2 \p_{\mu}\, \gamma_2\right),
\end{equation}
where $\varphi$, $\beta_i$ and $\gamma_i$ are functions of $A$ and $B$. We impose the boundary conditions $\varphi(0,0)=1$ and $\beta_i(0,0)$,
$\gamma_i(0,0)$ finite in order to ensure the smooth limit $\x_{\mu}\to x_{\mu}$ as $a\to 0$. Substituting the Ansatz \eqref{3.1} into
Eq. \eqref{2.1}, we obtain the following system of equations:
\begin{align}
\frac{\p\varphi}{\p A}\, \varphi-B \left(\frac{\p\varphi}{\p A} -2A\, \frac{\p\varphi}{\p B}\right)\beta_{2}+ \left(A\,
\frac{\p\varphi}
{\p A}+2B\frac{\p\varphi}{\p B}\right)\gamma_{1}-\varphi(\gamma_{1}-1) &=0, \label{3.2} \\
2\frac{\p\varphi}{\p B}\varphi-\left(\frac{\p\varphi}{\p A}-2A\frac{\p\varphi} {\p
B}\right)\beta_{1}-\left(A\frac{\p\varphi}{\p A}+2B\frac{\p\varphi} {\p B}\right)\gamma_{2}+\varphi\gamma_{2} &=0, \label{3.3}
\end{align}
\begin{align}
\Big(\frac{\p\beta_{1}}{\p A} -2B\frac{\p\beta_{2}}{\p B}\Big)\varphi -B\Big(\frac{\p\beta_{1}}{\p A}
-2A\frac{\p\beta_{1}}{\p B}\Big)\beta_{2}+ B\Big(\frac{\p\beta_{2}}{\p A}
-2A\frac{\p\beta_{2}}{\p B}\Big)\beta_{1} \notag \\
+\Big(A\frac{\p\beta_{1}}{\p A} +2B\frac{\p\beta_{2}}{\p B}\Big)\gamma_{1} +B\Big(A\frac{\p\beta_{2}}{\p A}
+2B\frac{\p\beta_{2}}{\p B}\Big)\gamma_{2} \notag \\
-\Big(\beta_{1}^2+2A\beta_{1}\beta_{2}\Big) +B\beta_{2}\gamma_{2}-2\beta_{2}\varphi+\beta_{1} &=0, \label{3.4} \\
-\Big(2\frac{\p\gamma_{1}}{\p B}+\frac{\p\gamma_{2}} {\p A}\Big)\varphi+\Big(\frac{\p\gamma_{1}} {\p
A}-2A\frac{\p\gamma_{1}}{\p B}\Big)\beta_{1}
+\Big(\frac{\p\gamma_{2}}{\p A}-2A\frac{\p\gamma_{2}}{\p B}\Big)B\beta_{2} \notag \\
+\Big(A\frac{\p\gamma_{1}}{\p A}+2B\frac{\p\gamma_{1}}{\p B}\Big)\gamma_{2} -\Big(A\frac{\p\gamma_{2}}{\p
A}+2B\frac{\p\gamma_{2}}{\p B}\Big)\gamma_{1} +\gamma_{2}(\beta_{1}-\gamma_{1}-1) &=0.  \label{3.5}
\end{align}

Next, we consider realizations of the rotation algebra $SO_a(n)$. We assume that the rotation generators are given by
the Ansatz
\begin{equation}\label{2.18}
M_{\mu\nu} = (x_{\mu}\p_{\nu}-x_{\nu}\p_{\mu})\, \F_1+i(x\p) (a_{\mu}\p_{\nu}-a_{\nu}\p_{\mu})\,
\F_2+i(x_{\mu}a_{\nu}-x_{\nu}a_{\mu}) \p^2\, \F_3,
\end{equation}
where the functions $\F_1$, $\F_2$ and $\F_3$ depend on $A$ and $B$, and satisfy the boundary conditions $\F_1(0,0)=1$ and
$F_2(0,0)$, $\F_3(0,0)$ finite, respectively. The boundary conditions imply that $M_{\mu\nu}$ becomes the ordinary rotation generator
as $a\to 0$. The Ansatz is antisymmetric in the indices $\mu$ and $\nu$. Now we can calculate the commutator $[M_{\mu\nu},M_{\lambda\rho}]$
by inserting the Ansatz \eqref{2.18} into Eq. \eqref{2.3B}. This results in the system of equations
\begin{equation}\label{F123}
\F_1 = 1, \qquad  \frac{\p \F_3}{\p A}+\left(\F_3+A\, \frac{\p \F_3}{\p A}\right) \F_2-2\F_3^2 = 0.
\end{equation}
Since $\F_2$ is uniquely determined by $\F_3$, these equations provide a realization of the algebra $SO_a(n)$ in terms of an
arbitrary parameter function $\F_3$. However, $M_{\mu\nu}$ and $\x_{\mu}$ are required to form the extended Lie
algebra \eqref{2.15}, hence $\F_3$ is related with the functions $\varphi$, $\beta_i$ and $\gamma_i$. This relation can be shown either
directly from Eq. \eqref{2.15} or by using a realization of the Dirac operator $D_{\mu}$ and Eq. \eqref{2.17}.

We assume that the Dirac operator is given by
\begin{equation}\label{2.16}
D_{\mu}=\p_{\mu}\, \G_1+ia_{\mu}\p^2\,  \G_2.
\end{equation}
Here the functions $\G_1$ and $\G_2$ depend on $A$ and $B$, and satisfy the boundary conditions $\G_1(0,0)=1$ and $\G_2(0,0)$ finite, respectively.
Inserting Eq. \eqref{2.16} into Eq. \eqref{2.4}, we obtain
\begin{align}
\sqrt{1-a^2 D^2}-iaD &= \G_1\varphi, \label{D-1} \\
\frac{\partial \G_1}{\partial A}\varphi+\Big(-\frac{\partial \G_1}{\partial A}+2A\frac{\partial \G_1}{\partial B}\Big)
B\beta_2+\Big(\G_1+A\frac{\partial \G_1}{\partial A}+2B\frac{\partial \G_1}{\partial B}\Big)\gamma_1 &= 0, \label{D-2} \\
2\frac{\partial \G_1}{\partial B}\varphi+\Big(2A\frac{\partial \G_1}{\partial B}-\frac{\partial \G_1}{\partial A}\Big)
\beta_1-\Big(\G_1+A\frac{\partial \G_1}{\partial A}+2B\frac{\partial \G_1}{\partial B}\Big)\gamma_2 &= 0, \label{D-3}
\end{align}
\begin{align}
\G_1 \beta_2+\frac{\partial \G_2}{\partial A}\varphi +2A\G_2\beta_2+B\Big(2A\frac{\partial \G_2}{\partial B}-
\frac{\partial \G_2}{\partial A}\Big)\beta_2  \notag \\
+\Big(2\G_2+A\frac{\partial \G_2}{\partial A}+2B\frac{\partial \G_2}{\partial B}\Big)\gamma_1 &=\G_2, \label{D-4} \\
\G_1\beta_1+\Big(2\G_2+2B\frac{\partial \G_2}{\partial B}\Big)\varphi+2A\G_2 \beta_1
+\Big(2A\frac{\partial \G_2}{\partial B}-\frac{\partial \G_2}{\partial A}\Big) B\beta_1  \notag  \\
-\Big(2\G_2+\frac{\partial \G_2}{\partial A}A+2B\frac{\partial \G_2}{\partial B}\Big) B\gamma_2 &=\G_1.  \label{D-5}
\end{align}
Equations \eqref{D-1} and \eqref{2.12} imply that the inverse shift operator has a simple realization as
\begin{equation}\label{Zinv}
Z^{-1}=\G_1\, \varphi.
\end{equation}

The system of PDE's for the unknown functions $\varphi$, $\beta_i$, $\gamma_i$, $\F_i$ and $\G_i$ is too difficult to
solve in full generality. We will reduce the system to a manageable form by considering special choices for the
functions $\beta_1$ and $\beta_2$: $\beta_1=\beta_2=0$ and $\beta_1=1$, $\beta_2=0$. In each case, we obtain an infinite family
of realizations parametrized by the function $\varphi$. We call these realizations type I and type II, respectively. \\
\\
\\

\noindent\textbf{Realization I:} $\beta_1=\beta_2=0$.

\noindent In the first realization, Eqs. \eqref{3.1}-\eqref{3.5} imply
\begin{equation}\label{RI.1}
\x_{\mu}=x_{\mu}\, \varphi+i(x\p) \left(a_{\mu}\gamma_1+ia^2 \p_{\mu}\, \gamma_2\right),
\end{equation}
where
\begin{align}
\gamma_1 &= \frac{\left(1+\frac{\p \varphi}{\p A}\right) \varphi}{\varphi-\left(A\, \frac{\p \varphi}{\p A}+2B\,
\frac{\p \varphi}{\p B}\right)}, \label{RI.2} \\
\gamma_2 &= -\frac{2\frac{\p \varphi}{\p B}\, \varphi}{\varphi-\left(A\, \frac{\p \varphi}{\p A}+2B\, \frac{\p
\varphi}{\p B}\right)}.  \label{RI.3}
\end{align}
Furthermore, Eqs. \eqref{D-2}-\eqref{D-5} yield
\begin{equation}\label{RI.4}
\G_1=\frac{1}{\varphi+A}, \quad \G_2 = \frac{1}{2\varphi\, (\varphi+A)}.
\end{equation}
We note that in view of Eq. \eqref{Zinv} the shift operator is given by
\begin{equation}\label{RI.5}
Z=1+\frac{A}{\varphi}.
\end{equation}
We are now able to find realizations of the rotation generators $M_{\mu\nu}$. Inserting the realizations for $\x_{\mu}$, $D_{\mu}$
and $Z$ into Eq. \eqref{2.17} and comparing the obtained expression with Eq. \eqref{2.18}, we find
\begin{equation}\label{RI.6}
\F_1 =1, \quad
\F_2 = \frac{\gamma_1}{\varphi}+\frac{B\, \gamma_2}{2\varphi^2}=\frac{\left(1+\frac{\p \varphi}{\p A}\right)\, \varphi-B\,
\frac{\p\varphi}{\p B}}{\varphi^2-\left(A\frac{\p \varphi}{\p A}+2B\, \frac{\p \varphi}{\p B}\right)\, \varphi}, \quad
\F_3 = \frac{1}{2\varphi}.
\end{equation}
Note that $\F_1=1$ is consistent with the earlier obtained expression in Eq. \eqref{F123}. Next, we find a realization
of the Laplace operator $\square$ which is uniquely determined by Eq. \eqref{2.7}. Using the realizations for $Z$ and $D_{\mu}$ in 
Eqs. \eqref{RI.4} and \eqref{RI.5} we obtain $\sqrt{1-a^2 D^2} = 1-a^2 \p^2 \G_2$. Therefore, Eq. \eqref{2.7} yields
\begin{equation}\label{RI.7}
\square = \p^2 \H ,\quad \text{where} \quad \H = \frac{1}{\varphi (\varphi+A)}.
\end{equation}
Note that since $\varphi(0,0)=1$ and $A\to 0$ as $a\to 0$, in the limiting case we have $\square \to \p^2$ as $a\to 0$.\\

\noindent\textbf{Realization II:} $\beta_1=1$, $\beta_2=0$.\\
\noindent Repeating the above calculations in the second realization we find
\begin{equation}
\x_{\mu} = x_{\mu}\, \varphi+i(ax)\p_{\mu}+i(a\p) \left(a_{\mu}\gamma_1+ia^2\p_{\mu}\, \gamma_2\right),
\end{equation}
where
\begin{align}
\gamma_1 &= \frac{\left(1+\frac{\p \varphi}{\p A}\right)\, \varphi}{\varphi-\left(A\, \frac{\p \varphi}{\p A}+2B\,
\frac{\p \varphi}{\p B}\right)}, \\
\gamma_2 &= \frac{\frac{\p \varphi}{\p A}-2(\varphi+A)\frac{\p \varphi}{\p B}}{\varphi-\left(A\, \frac{\p \varphi}{\p
A}+2B\, \frac{\p \varphi}{\p B}\right)}.
\end{align}
The functions $\G_1$ and $\G_2$ are given by
\begin{equation}\label{2.19}
\G_1=\frac{1}{\sqrt{(\varphi+A)^2+B}}, \quad \G_2=0,
\end{equation}
and the shift operator yields
\begin{equation}
Z=\sqrt{\left(1+\frac{A}{\varphi}\right)^2+\frac{B}{\varphi^2}}.
\end{equation}
Similarly, for $\F_1$, $\F_2$ and $\F_3$ we obtain
\begin{equation}
\F_1 =1, \quad
\F_2 = \frac{\gamma_1}{\varphi} = \frac{1+\frac{\p \varphi}{\p A}}{\varphi-\left(A\, \frac{\p \varphi}{\p A}+2B\, \frac{\p \varphi}{\p B}\right)},
\quad \F_3 = 0.
\end{equation}
Note that $\F_1$, $\F_2$ and $\F_3$ in all realizations are consistent with Eq. \eqref{2.17}.
Furthermore, for the Laplace operator we have $\square = \p^2 \H$ where
\begin{equation}
\H = \frac{2}{B}\left(1-(\varphi+A)\, \G_1\right) = \frac{2}{B} \left(1-\frac{\varphi+A}{\sqrt{(\varphi+A)^2+B}}\right).
\end{equation}

\subsection{Special realizations}

Of particular interest are some realizations obtained for a special choice of the parameter function $\varphi$. In the
realization of type I we consider the \textit{left}, \textit{right} and \textit{symmetric} realizations corresponding to $\varphi_L=1-A$,
$\varphi_R=1$ and $\varphi_S=A/(\exp(A)-1)$, respectively. One can show that in the left realization the derivative operator $\p_{\mu}$ 
becomes the left deformation derivative $\p_{\mu}^L$. Similarly, in the right realization the derivative operator $\p_{\mu}$ becomes 
the right deformation derivative $\p_{\mu}^R$. The symmetric realization is related to Weyl's symmetric ordering of the monomials in $\x_{\mu}$.
In the realization of type II we consider the \textit{natural} realization corresponding to $\varphi_N=-A+\sqrt{1-B}$. In this realization
the Dirac derivative is simply given by $D_{\mu}=\p_{\mu}$. \\

\noindent \textbf{Left realization:} $\beta_1=\beta_2=0$, $\varphi_L=1-A$. \\
Inserting $\varphi_L=1-A$ into Eqs. \eqref{RI.1}-\eqref{RI.7}, we find
\begin{align}
\hat x_{\mu} &= x_{\mu}(1-A), \label{left.1} \\
M_{\mu\nu} &= x_{\mu}\partial_{\nu}-x_{\nu}\partial_{\mu}+\frac{1}{2}i(x_{\mu}a_{\nu}-x_{\nu}a_{\mu})\frac{1}{1-A}\partial^2, \label{left.2} \\
D_{\mu} &= \partial_{\mu}+\frac{ia_{\mu}}{2} \square, \label{left.3} \\
Z &= \frac{1}{1-A}, \label{left.4} \\
\square &= \frac{1}{1-A}\partial^2.  \label{left.5}
\end{align}
It follows form Eqs. \eqref{left.1} and \eqref{left.4} that
\begin{equation}
[\p_{\mu},\x_{\nu}] = \delta_{\mu\nu} Z^{-1}.
\end{equation}
Thus, in view of Eq. \eqref{2.10B}, we see that $\p_{\mu}$ is the left deformation derivative $\p_{\mu}^L$. \\

\noindent \textbf{Right realization:} $\beta_1=\beta_2=0$, $\varphi_R=1$. \\
Repeating the calculations with $\varphi_R=1$, we obtain
\begin{align}
\x_{\mu} &= x_{\mu}+ia_{\mu} (x\p), \\
M_{\mu\nu} &= x_{\mu} \p_{\nu}-x_{\nu} \p_{\mu} + i(x\p) (a_{\mu}\p_{\nu}-a_{\nu}\p_{\mu})+\frac{i}{2}(x_{\mu}a_{\nu}-a_{\nu}x_{\mu})\p^2, \\
D_{\mu} &= \frac{1}{1+A}\p_{\mu}+\frac{ia_{\mu}}{2}\square, \\
Z &= 1+A, \\
\square &= \frac{1}{1+A}\p^2.
\end{align}
In this case,
\begin{equation}
[\p_{\mu},\x_{\nu}] = \delta_{\mu\nu}+ia_{\nu} \p_{\mu},
\end{equation}
hence a comparison with Eq. \eqref{2.11B} shows that $\p_{\mu}$ is the right deformation derivative $\p_{\mu}^R$. \\

\noindent \textbf{Symmetric realization:} $\beta_1=\beta_2=0$, $\varphi_S=A/(\exp(A)-1)$. \\
This realization corresponds to the symmetric Weyl ordering \cite{22}. It also follows from the universal formula for a general Lie algebra \cite{41}, 
after inserting  the structure constants from Eq. \eqref{2.2}.
 We have
\begin{align}
\hat x_{\mu} &= x_{\mu} \frac{A}{e^A-1}+ia_{\mu}(x\partial) \frac{e^A-1-A}{(e^A-1)\, A}, \\
M_{\mu\nu} &= x_{\mu}\partial_{\nu}-x_{\nu}\partial_{\mu}+i(x\partial) (a_{\mu}\partial_{\nu}-a_{\nu}\partial_{\mu})\frac{e^A-1-A}{A^2}, \\
D_{\mu} &= \frac{e^A-1}{A\, e^A}\partial_{\mu}+\frac{ia_{\mu}}{2}\square, \\
Z &= e^A, \\
\square &= \frac{(e^A-1)^2}{A^2\, e^A} \partial^2, \\
\intertext{and}
[\p_{\mu},\x_{\nu}] &= \delta_{\mu\nu}\varphi_S+ia_{\nu} \p_{\mu}\frac{1-\varphi_S}{A}.
\end{align}

\noindent\textbf{Natural realization:} $\beta_1=1$, $\beta_2=0$, $\varphi_N=-A+\sqrt{1-B}$. \\
The natural realization is a special case of the realization of type II in which the Dirac derivative $D_{\mu}=\p_{\mu} \G_1+ia_{\mu} \p^2 \G_2$
simplifies to $D_{\mu}=\p_{\mu}$. In view of Eq. \eqref{2.19}, this holds when $\varphi_N=-A+\sqrt{1-B}$. Then we have
\begin{align}
\x_{\mu} &= x_{\mu} \left(-A+\sqrt{1-B}\right)+i(ax)\, \p_{\mu}, \label{2.20} \\
M_{\mu\nu} &= x_{\mu} \p_{\nu}-x_{\nu} \p_{\mu}, \label{2.21} \\
D_{\mu} &= \p_{\mu},  \label{2.22} \\
Z &= \frac{1}{-A+\sqrt{1-B}}, \\
\square &= \frac{2}{B}\left(1-\sqrt{1-B}\right) \p^2.
\end{align}
In the natural realization it is easily shown that the rotation generators $M_{\mu\nu}$ are given by Eq. \eqref{2.17}.
Indeed, Eqs. \eqref{2.20} and \eqref{2.22} imply that
\begin{equation}\label{add-2}
(\x_{\mu}D_{\nu}-\x_{\nu}D_{\mu})Z = (x_{\mu}\p_{\nu}-x_{\nu}\p_{\mu})\, \varphi Z.
\end{equation}
However, $\varphi Z=1$, thus Eqs. \eqref{2.21} and \eqref{add-2} yield
\begin{equation}
M_{\mu\nu}=(\x_{\mu} D_{\nu}-\x_{\nu} D_{\mu}) Z.
\end{equation}

\subsection{Hermiticity}

All relations of the type $[\x_{\mu},\x_{\nu}]$, $[M_{\mu\nu},M_{\lambda\rho}]$, $[M_{\mu\nu},\x_{\lambda}]$, $[D_{\mu},D_{\nu}]$,
$[M_{\mu\nu},D_{\lambda}]$, $[D_{\mu},\x_{\nu}]$, i.e. Eqs. \eqref{2.1}, \eqref{2.3B}, \eqref{2.15}, \eqref{2.24}, \eqref{2.25}
and \eqref{2.4}, are invariant under the formal antilinear involution:
\begin{equation}
\hat{x}_\mu^{\dag}=\hat{x}_\mu,\quad D_\mu^{\dag}=-D_\mu , \quad
M_{\mu\nu}^{\dag}=-M_{\mu\nu},\quad c^{\dag}=\bar{c}, \quad c\in \C.
\end{equation}
The order of elements in the product is inverted under the involution. The commutative coordinates $x_\mu$ and their
derivatives $\partial_\mu$ also satisfy the involution property: $x_\mu^{\dag}=x_\mu$ and $\partial_\mu^{\dag}=-\partial_\mu$.
Then the NC coordinates $ \hat{x}_\mu$ are represented by hermitian operators. However, Eq. \eqref{3.6} is generally
\textit{not} hermitian. The hermitian representations are simply obtained by the following expression \cite{22}:
\begin{equation}
\hat{x}_{\mu}^h=\frac{1}{2}\Big(x_{\alpha} \Phi_{\alpha\mu}+(\Phi^{\dag})_{\mu\alpha}x_{\alpha}\Big).
\end{equation}
However, the physical results do not depend on the choice of representation as long as there exists a smooth limit
$\x_{\mu}\to x_{\mu}$ as $a \to 0$. Therefore, we restrict ourselves to non-hermitian realizations only.

\section{Star product}

Recall that in Sect. \ref{covariant-realizations} a general Ansatz for the NC coordinates was introduced,
\begin{equation}\label{SP.1}
\x_{\mu} = x_{\alpha} \Phi_{\alpha\mu}(A,B), \quad \Phi_{\alpha\mu}(0,0)=\delta_{\alpha\mu}.
\end{equation}
Let us define the vacuum state by $|0\rangle = 1$ and $\p_{\mu} |0\rangle =0$, and fix the normalization condition by
$\x_{\mu}|0\rangle =x_{\mu}$. For a given realization $\Phi_{\mu\nu}$ there is a unique map sending mononials in the NC coordinates
$\x_{\mu}$ into polynomials of the commutative coordinates $x_{\mu}$. This map is given by
\begin{equation}\label{SP.2}
\prod_{i=1}^k \x_{\mu_i} |0\rangle = P_k(x),
\end{equation}
where $P_k$ is a polynomial of degree $k$. We also have the dual relation
\begin{equation}\label{SP.3}
\prod_{i=1}^k x_{\mu_i} = \hat P_k(\x)|0\rangle,
\end{equation}
where $\hat P_k$ is also a polynomial of degree $k$ in $\x$. For example, in the left realization we have
\begin{equation}
\prod_{i=1}^k \x_{\mu_i} |0\rangle = x_{\mu_1}(x_{\mu_2}-ia_{\mu_2})(x_{\mu_3}-i2a_{\mu_3})\ldots (x_{\mu_k}-i(k-1)a_{\mu_k}),
\end{equation}
together with the dual relation
\begin{equation}
\prod_{i=1}^k x_{\mu_i} = \x_{\mu_1}Z\x_{\mu_2}\ldots Z\x_{\mu_k} |0\rangle.
\end{equation}
Similarly, in the right realization we find
\begin{align}
\prod_{i=1}^k \x_{\mu_i} |0\rangle &= (x_{\mu_1}+i(k-1)a_{\mu_1})(x_{\mu_2}+i(k-2)a_{\mu_2})\ldots (x_{\mu_{k-1}}+ia_{\mu_{k-1}}) x_{\mu_k}, \\
\prod_{i=1}^k x_{\mu_i} &= Z^{-(k-1)}\x_{\mu_1} Z \x_{\mu_2}\ldots Z\x_{\mu_k}|0\rangle.
\end{align}
One can obtain similar expressions for the symmetric realization \cite{22}. It is interesting to note that
in the realization of type I when $\varphi = \varphi (A)$ the following relation holds:
\begin{equation}
e^{ik\x}|0\rangle = \exp\left(i\frac{\varphi(-ak)}{\varphi_S (-ak)}\, kx\right),
\end{equation}
where $k\in \mathbb{\R}^n$ and $\varphi_S(A)=A/(\exp(A)-1)$. In particular, in the symmetric realization when $\varphi=\varphi_S$ we have
\begin{equation}
e^{ik\hat{x}}|0\rangle = e^{ikx}.
\end{equation}
Similarly, in the natural realization one can show that
\begin{equation}\label{3.11}
e^{ik\x}|0\rangle = e^{iP_N(k) x},
\end{equation}
where
\begin{equation}\label{3.12}
P_N(k)_\mu=\frac{1}{\varphi_S (ak)}\left(k_\mu-\frac{k^2}{2\varphi_S (-ak)}\, a_\mu\right).
\end{equation}
Equation \eqref{SP.2} defines an isomorphism of the universal enveloping algebras generated by
$\x_1,\x_2,\ldots ,\x_n$ and $x_1,x_2,\ldots ,x_n$, respectively.  This can be extended to formal power series by
\begin{equation}
\hat f(\x) |0\rangle = f(x),
\end{equation}
where the function $f$ depends on the realization $\Phi_{\mu\nu}$. The Leibniz rule
and the related coproduct $\triangle \p_{\mu}$ follow uniquely from the commutator relation
\begin{equation}
[\p_{\mu},\x_{\nu}] = \Phi_{\mu\nu}(A,B)
\end{equation}
and the conditions $\p_{\mu}|0\rangle =0$, $\x_{\mu} |0\rangle = x_{\mu}$.

The star product of two functions $f(x)$ and $g(x)$ is defined by
\begin{equation}
(f\star g)(x) = \hat f(\x) \hat g(\x) |0\rangle.
\end{equation}
We emphasize that the star product depends on the realization $\Phi_{\mu\nu}$. The result relating the star product and the coproduct 
obtained for non-covariant realizations \cite{22} can be extended to covariant realizations of $\kappa$-deformed space, Eq. \eqref{2.1}, as
\begin{equation}\label{SP.4}
(f\star g)(u) = m\left(e^{u_{\alpha} (\triangle-\triangle_0) \p_{\alpha}} f(x)\otimes g(y)\right)\Big|_{\substack{x=u \\ y=u}},
\end{equation}
where $m$ is the multiplication map in the Hopf algebra and $\triangle_0 \p_{\mu} = \p_{\mu}\otimes \mathbf{1}+\mathbf{1}\otimes
\p_{\mu}$ is the undeformed coproduct.

\subsection{Star product for the realization of type I}

In this subsection we discuss the star product in the realization of type I assuming that the parameter function $\varphi$ depends
only on the variable $A=ia\p$. In view of Eq. \eqref{RI.1}, we have
\begin{equation}
\x_{\mu} = x_{\mu} \varphi(A)+i(x\p) a_{\mu} \gamma (A),
\end{equation}
where
\begin{equation}
\gamma(A) = \frac{1+\varphi^\prime (A)}{1-A\frac{\varphi^\prime (A)}{\varphi (A)}}, \quad \varphi (0) = 1.
\end{equation}
In order to find the coproduct $\triangle \p_{\mu}$, we note that in this realization $\p_{\mu} = \varphi(A) \p_{\mu}^R$.
The coproduct for $\p_{\mu}^R$ is given by Eq. \eqref{coproductR}, hence
\begin{align}
\triangle \p_{\mu} &= \triangle \varphi(A) \triangle \p_{\mu}^R \notag \\
&= \triangle \varphi(A) \left[\frac{1}{\varphi(A)} \p_{\mu}\otimes \mathbf{1}+Z\otimes \frac{1}{\varphi(A)} \p_{\mu}\right]. \label{3.7}
\end{align}
Inverting the expression for the shift operator $Z=1+A/\varphi(A)$, we find
\begin{equation}\label{SP.5}
A=(Z-1)+\varphi^\prime (0)(Z-1)^2 + \cdots ,
\end{equation}
which, together with $\triangle Z = Z\otimes Z$, allows us to calculate the coproduct $\triangle \varphi (A)$.
Then, to second order in the parameter $a$, Eq. \eqref{3.7} leads to
\begin{equation}
\begin{split}
\triangle \p_{\mu} &= \p_{\mu}^x \left[1+\varphi^{\prime}(0)
A_y+\left(\varphi^{\prime\prime}(0)+(\varphi^{\prime}(0))^2+\varphi^{\prime}(0)\right) A_x A_y+\frac{1}{2}
\varphi^{\prime\prime}(0) A_y^2\right] \\
&+\p_{\mu}^y\left[1+(1+\varphi^{\prime}(0))A_x+\left(\varphi^{\prime\prime}(0)+(\varphi^{\prime}(0))^2+
\varphi^{\prime}(0)\right) A_x A_y+\frac{1}{2} \varphi^{\prime\prime}(0) A_x^2\right],
\end{split}
\end{equation}
where $A_x = ia\p^x$, $A_y=ia\p^y$. Consequently, the star product from Eq. \eqref{SP.4} is given by
\begin{equation}
\begin{split}
(f \star g)(u) &=\Big\{1+u\p^x \Big[\varphi^{\prime}(0) A_y +
\left(\varphi^{\prime\prime}(0)+(\varphi^{\prime}(0))^2+\varphi^{\prime}(0)\right) A_x A_y+\frac{1}{2}
\varphi^{\prime\prime}(0) A_y^2\Big] \\
&+u\p_y\Big[(1+\varphi^{\prime}(0))A_x +
\left(\varphi^{\prime\prime}(0)+(\varphi^{\prime}(0))^2+\varphi^{\prime}(0)\right) A_x A_y +\frac{1}{2}
\varphi^{\prime\prime}(0) A_x\Big]  \\
&+ \frac{1}{2}\Big[\varphi^{\prime}(0)\, u \p^x A_y + (1+\varphi^{\prime}(0))\, u\p^y A_x\Big]^2\Big\} f(x)
g(y)\Big|_{\substack{x=u \\ y=u}}.  \label{3.8}
\end{split}
\end{equation}
One can show that the dual relation holds
\begin{equation}
(f\star g)_{\varphi(A)} = (g\star f)_{\varphi(-A)-A},
\end{equation}
where the star products correspond to the functions $\varphi(A)$ and $\varphi(-A)-A$, respectively.

\subsection{Star product for special realizations}

In this subsection we give the star products in closed form for the left, right and symmetric realizations, as well
as the star product to second order in $a$ in the natural realization. \\

\noindent \textbf{Left realization:} $\varphi_L = 1-A$.
\begin{equation}
(f\star g)_{\varphi_L}(u)=e^{-u_{\alpha} \p_{\alpha}^x A_y} f(x) g(y)\Big|_{\substack{x=u\\ y=u}}.
\end{equation}

\noindent \textbf{Right realization:} $\varphi_R =1$.
\begin{equation}\label{SPright}
(f\star g)_{\varphi_R}(u) = e^{u_{\alpha} \p_{\alpha}^y A_x} f(x) g(y)\Big|_{\substack{x=u\\ y=u}}.
\end{equation}
The ``left'' and ``right'' star products satisfy the symmetry relation
\begin{equation}
(f\star g)_{\varphi_L} =(g \star f)_{\varphi_R}.
\end{equation}

\noindent \textbf{Symmetric realization:} $\varphi_S = A/(\exp(A)-1)$.
\begin{equation}\label{SP.19}
(f\star g)_{\varphi_S}(u) = e^{u_{\alpha} (\triangle-\triangle_0) \p_{\alpha}} f(x) g(y)\Big|_{\substack{x=u\\ y=u}},
\end{equation}
where
\begin{align}
\triangle_0 \p_{\mu} &= \p_{\mu}^x + \p_{\mu}^y, \\
\triangle \p_{\mu} &= \p_{\mu}^x\, \frac{\varphi_S(A_x+A_y)}{\varphi_S (A_x)}+\p_{\mu}^y\, \frac{\varphi_S (-A_x-A_y)}{\varphi_S (-A_y)}. \label{SP.20}
\end{align}
In this case, we have
\begin{equation}
(f\star g)_{\varphi_{S} (A)} = (g\star f)_{\varphi_{S} (-A)}.
\end{equation}
\vskip 0.5cm
The symmetric realization $\varphi =\varphi_S$ corresponds to the symmetric Weyl ordering \cite{22}. Our closed form results, Eqs. \eqref{SP.19} 
and \eqref{SP.20}, are in agreement with the general series expansion formula for the star product of the Lie algebra type  NC space \cite{42}.\\

\noindent \textbf{Natural realization:} $\varphi_N = -A+\sqrt{1-B}$. \\
In this realization we have $\p_{\mu}=D_{\mu}$, hence the coproduct $\triangle \p_{\mu}$ is given
by Eq. \eqref{coproductD}. One can show that to second order in the parameter $a$ the star product yields
\begin{equation}
\begin{split}
(f\star g)_{\varphi_N}(u) &= f(u) g(u)   \\
&+\Bigg\{u_{\mu} \Big[\Big(-\p_{\mu}^x-\frac{ia_{\mu}}{2a^2}\, a_{\alpha}^2\, (\p_{\alpha}^x)^2\Big) A_y
-\frac{1}{2}\p_{\mu}^x\, a_{\alpha}^2\, (\p_{\alpha}^y)^2+ia_{\mu} (1+A_x)\p^x \p^y \Big]   \\
&+\frac{1}{2} u_{\mu} u_{\nu}\Big[\p_{\mu}^x \p_{\nu}^y A_y^2-2i a_{\mu} \p_{\nu}^x\, \p^x \p^y\, A_y-a_{\mu} a_{\nu} (\p^x \p^y)^2\Big]\Bigg\}
f(x) g(y)\Big|_{\substack{x=u\\ y=u}},
\end{split}
\end{equation}
where $\p^x \p^y = \p_{\alpha}^x\, \p_{\alpha}^y$.

\subsection{Equivalent star products}

So far we have considered the star product for some specific realizations of type I and II. We point out, however, that
infinitely many realizations of the star product can be constructed by similarity transformations of the variables $x_{\mu}$, $\p_{\mu}$.
In the following we consider the star products obtained by similarity transformations starting
from the right realization.

Recall that in the right realization we have
\begin{equation}\label{SP.6}
\x_{\mu} = x_{\mu}^R+ia_{\mu} (x^R\, \p^R),
\end{equation}
and the star product is given by Eq. \eqref{SPright}. (From now on the variables $x_{\mu}$ and $\p_{\mu}$ used in the right realization
will be denoted by $x_{\mu}^R$ and $\p_{\mu}^R$, respectively.) A similarity transformation is defined by
\begin{align}
x_{\mu} &= S^{-1} x_{\mu}^R S, \label{SP.8} \\
\p_{\mu} &= S^{-1} \p_{\mu}^R S. \label{SP.9}
\end{align}
Clearly, the new variables $x_{\mu}$ and $\p_{\mu}$ also generate the Heisenberg algebra. We define the vacuum condition on $S$
by $S|0\rangle = |0\rangle$. In view of Eq. \eqref{SPright}, the star product induced by the similarity transformation $S$ is given by
\begin{align}
(f\underset{S}{\star} g)(u) &= S\left(S^{-1}f \star S^{-1}g\right)_{\varphi_R} (u) \notag \\
&= S e^{u_{\alpha} \p_{\alpha}^y A_x}\, (S^{-1}f)(x)\, (S^{-1}g)(y)\Big|_{\substack{x=u\\ y=u}}.  \label{3.9}
\end{align}
Two star products are said to be equivalent if they are related by a similarity transformation. For example, if $S=e^{-x\p\, A_x}$, then
\begin{equation}
(f\star g)_{\varphi_L}(u) = S\left(S^{-1} f\star S^{-1}g\right)_{\varphi_R} (u),
\end{equation}
hence the star products for the left and right realization are equivalent. We will show that all realizations of type I with
$\varphi = \varphi (A)$ lie in the orbits of the action of similarity transformations of the right realization. Hence, any
two star products in realizations of this type are equivalent.

Consider a realization of type I,
\begin{equation}\label{SP.7}
\x_{\mu} = x_{\mu}\varphi (A)+ia_{\mu} (x\p)\frac{1+\varphi^\prime (A)}{1-A\frac{\varphi^\prime (A)}{\varphi (A)}}, \quad \varphi (0)=1.
\end{equation}
The transformation $(x_{\mu},\p_{\mu}) \mapsto (x_{\mu}^R, \p_{\mu}^R)$ which maps the realization \eqref{SP.6} into
\eqref{SP.7} is given by
\begin{align}
x_{\mu}^R &= x_{\mu} \varphi (A)+ia_{\mu} (x\p) \frac{\varphi^\prime (A)}{1-A\frac{\varphi^\prime (A)}{\varphi (A)}},  \label{SP.10}\\
\p_{\mu}^R &= \p_{\mu}\, \frac{1}{\varphi (A)}. \label{SP.11}
\end{align}
We show that there exists a similarity operator of the form $S=\exp(U)$,
\begin{equation}\label{3.10}
U=(x\p) \sum_{k=1}^\infty c_k A^k,
\end{equation}
such that Eqs. \eqref{SP.10}-\eqref{SP.11} are given by $x_{\mu}^R=S x_{\mu} S^{-1}$ and $\p_{\mu}^R = S \p_{\mu} S^{-1}$, respectively.
Then, Eq. \eqref{SP.11} yields
\begin{equation}\label{SP.12}
\exp(\text{ad}(U)) \p_{\mu} = \p_{\mu} \frac{1}{\varphi (A)}.
\end{equation}
By expanding both sides of Eq. \eqref{SP.12} into power series in $A$, one can show that the coefficients $c_k$ are uniquely determined
by the function $\varphi (A)$. Expanding the right-hand side of Eq. \eqref{SP.12} leads to
\begin{equation}\label{SP.13}
\p_{\mu} \frac{1}{\varphi (A)} = \p_{\mu}\left(1+\sum_{p=1}^\infty \frac{1}{p!}\, \varepsilon_p\, A^p\right),
\end{equation}
where the coeffcients $\varepsilon_p$ can be found recursively from
\begin{align}
\varepsilon_1 &= -\varphi^\prime (0), \label{SP.17} \\
\varepsilon_p &= -\varphi^{(p)}(0)-\sum_{k=1}^{p-1} \binom{p}{k} \varphi^{(k)}(0)\, \varepsilon_{p-k}, \quad p\geq 2. \label{SP.18}
\end{align}
Similarly, the expansion of the left-hand side of Eq. \eqref{SP.12} gives
\begin{equation}\label{SP.14}
\exp(\text{ad}(U)) \p_{\mu} = \p_{\mu}\left(1+\sum_{p=1}^\infty \gamma_p\, A^p\right),
\end{equation}
where the coefficients $\gamma_p$ are found in terms of $c_k$ as follows:
\begin{align}
\gamma_1 &= -c_1, \\
\gamma_p &= -c_p+\sum_{n=2}^p \frac{(-1)^n}{n!}\, \beta_p^{(n)}, \quad p\geq 2.
\end{align}
Here,
\begin{equation}
\beta_p^{(n)} = \sum_{|k|=p} \Psi(k)\, \prod_{i=1}^n c_{k_i},
\end{equation}
where $k=(k_1,k_2,\ldots ,k_n)$, $k_i\in \mathbb{N}$, is a multi-index with length $|k|=\sum_{i=1}^n k_i$, the function $\Psi(k)$
is defined by
\begin{equation}
\Psi(k)=(1+k_n)(1+k_n+k_{n-1})\ldots (1+\sum_{k=2}^n k_i),
\end{equation}
and the summation is taken over all multi-indices such that $|k|=p$. It follows from Eqs. \eqref{SP.13} and \eqref{SP.14} that
\begin{align}
c_1 &= -\varepsilon_1, \label{SP.15} \\
c_p &= \sum_{n=2}^p \frac{(-1)^n}{n!}\, \beta_p^{(n)}-\frac{1}{p!}\, \varepsilon_p, \quad p\geq 2.  \label{SP.16}
\end{align}
Note that the coefficient $c_p$ is uniquely determined by $c_1,c_2,\ldots ,c_{p-1}$. Hence, Eqs. \eqref{SP.15}-\eqref{SP.16},
together with Eqs. \eqref{SP.17}-\eqref{SP.18}, provide recursion relations for $c_p$ in terms of $\varphi^{(k)}(0)$,
thus $S$ is uniquely determined by $\varphi (A)$. For example, the first
three coefficients are given by $c_1=\varphi^{\prime}(0)$, $c_2=\frac{1}{2}\varphi^{\prime\prime}(0)$,
$c_3=\frac{1}{6}\varphi^{\prime\prime\prime}(0)+\frac{1}{4}\varphi^{\prime}(0)\, \varphi^{\prime\prime}(0)$.
Equations \eqref{3.9} and \eqref{3.10}, together with $c_1$ and $c_2$, reproduce the star product given by Eq. \eqref{3.8}. 
This represents an important consistency check of our approach.

Note that our covariant realizations for $\hat{x}_\mu$, Eq. \eqref{3.1}, and for $D_\mu$, Eq. \eqref{2.16}, follow from $S=e^U$, where
\begin{equation}
U=(x\p)\Phi_1(A,B)+(xa)\p^2\Phi_2(A,B),
\end{equation}
with boundary conditions $\Phi_1(0,0)=0$ and $\Phi_2(0,0)$ finite.

\subsection{Scalar fields and invariants on kappa-deformed space}

All covariant realizations are physically equivalent. Here we consider the natural realization $x^N$, $\p^N$ defined by
$D_{\mu}=\p^N_{\mu}$ and $M_{\mu\nu}=x^N_{\mu} \p^N_{\nu}-x^N_{\nu} \p^N_{\mu}$ (c.f. Sects. 3.1 and 4.2). The realization
of the NC coordinates is given by Eq. \eqref{2.20},
\begin{equation}
\x_{\mu} = x^N_{\mu} Z^{-1}+i(ax^N) \p^N_{\mu}.
\end{equation}
Let us consider a scalar field $\hat \Phi (\x)$ in the NC coordinates satisfying $[M_{\mu\nu},\hat \Phi (\x)]=0$ for all
$\mu$, $\nu$. Then we define a scalar field $\Phi (x^N)$ in the undeformed space by
\begin{equation}
\hat \Phi \left(\x (x^N)\right) |0\rangle = \Phi (x^N).
\end{equation}
The ordinary Fourier transform is defined by
\begin{equation}
\tilde \Phi (k) = \int d^n x^N\, e^{-ik x^N} \Phi (x^N), \quad k\in \R^n.
\end{equation}
Then using the relation \eqref{3.11},
\begin{equation}
e^{ik \x} |0\rangle = e^{iP_N(k) x^N},
\end{equation}
where $P_N(k)$ is given by Eq. \eqref{3.12}, we find
\begin{equation}\label{3.13}
\hat \Phi (\x) = \int d^n k\, \tilde \Phi (k) e^{iP^{-1}_N(k) \x}
\end{equation}
which holds in any realization of $\hat{x}$. Here $P_N^{-1}$ denotes the inverse function of $P_N$,
\begin{equation}
P_N^{-1}(k)_\mu=\frac{\ln Z^{-1}(k)}{Z^{-1}(k)-1}\Big(k_\mu+\frac{a_\mu}{a^2}(\sqrt{1+a^2\, k^2}-1)\Big),
\end{equation}
where
\begin{equation}
Z^{-1}(k)=\sqrt{1+a^2\, k^2}+ak.
\end{equation}
The above relation \eqref{3.13} represents  a construction of $SO_a(n)$ invariants $\hat \Phi (\x)$ in terms of $\Phi (x^N)$ and 
$\tilde{\Phi} (k)$ by using the natural realization \eqref{2.20}. Alternatively, from Eq. \eqref{2.20} we obtain the inverse mapping
\begin{equation}
x^N_{\mu} = \Big(\x_{\mu}-i(a\x) \frac{\p^N_{\mu}}{\sqrt{1-a^2 (\p^N_{\mu})^2}}\Big) Z.
\end{equation}
Then, we find
\begin{equation}
\Phi \left(x^N (\x)\right) |0\rangle = \hat \Phi (\x).
\end{equation}
Both constructions are consistent and equivalent. Furthermore, we define the invariant integration over the entire NC space using
the natural realization as follows:
\begin{equation}
\int \hat \Phi (\x) = \int d^n x^N\, \Phi (x^N)
\end{equation}
with the property
\begin{equation}
\int \hat \Phi_1(\x) \hat \Phi_2 (\x) = \int d^n x^N\,(\Phi_1 \star \Phi_2)_N (x^N).
\end{equation}
 A generalized action of the scalar field $\Phi (\x)$ on the NC $\kappa$-deformed space is simply the action of the ordinary scalar field
in natural coordinates $x^N$ on the undeformed space in which pointwise multiplication of fields is replazed by $\star$-mulitplication
in the natural realization. Further inverstigation of this problem is in progress and will be published separately.

\section{Conclusion}

We have constructed covariant realizations of a general $\kappa$-deformed space in terms of commutative coordinates
$x_\mu$ and their derivatives $\p_\mu$ in the undeformed space. Our construction can also be applied to spaces
with arbitrary signatures, especially to Minkowski-type spaces.

Particulary, we have studied $\kappa$-deformed space whose deformation is described by an arbitrary vector. The NC coordinates
and rotation generators form an extended Lie algebra. The subalgebra of rotation generators, $SO_a(n)$,
is undeformed. The Dirac derivatives mutually commute and are vector-like under the action of $SO_a(n)$.
By introducing the shift operator the deformed Heisenberg algebra is written in a very simple way.
We have presented the Leibniz rule and coproduct for the rotation generators $M_{\mu\nu}$ and
the Dirac derivatives $D_\mu$ in a covariant form.

We have found two types of covariant realizations which are described by an arbitrary function $\varphi (A)$ with
$\varphi (0)=1$. We point out that all covariant realizations are equivalent and on an equal footing.
We have constructed coproducts and star products for covariant realizations. There is an important relation
between the coproduct and the star product in terms of an exponential map for a given realization. Specially, we have
found a few realizations (covariant left, right, symmetric and natural) which have very simple properties.
All realizations of type I are related by similarity transformations and the corresponding star products
are equivalent. We have considered scalar fields in NC coordinates and demonstrated their simple
properties using the natural realization. Constructions of invariants and invariant integration on NC spaces are also discussed.

Our approach may be useful in quantum gravity models, specially in $2+1$ dimensions. In this case,
the coresponding Lie algebra is $SU(2)$ or $SU(1,1)$ \cite{43}, \cite{44}, \cite{45}. It would be interesting to classify
NC spaces with covariant realizations in which NC coordinates and rotation generators form an extended
Lie algebra. For example, Snyder space is of this type and covariant realizations in terms of undeformed space exist \cite{46}.\\

\noindent{\bf Acknowledgments}\\

One of us  (S.M.) thanks D. Svrtan  and  Z. {\v{S}}koda for useful
discussions. This work is supported by the Ministry of Science and Technology of
the Republic of Croatia under the contracts 098-0000000-2865 and  0023003.

\end{document}